\documentclass[
    ,final            
    ,numberedheadings 
  ]
  {aipproc}

\layoutstyle{6x9}

\begin{document}

\title{Evolution of event-by-event $E_T$ fluctuations over collision 
centrality in RHIC interactions}

\classification{25.75.-q}
\keywords      {ET, Transverse energy, Event-by-event, Fluctuations}

\author{Raul Armendariz \\
for the PHENIX Collaboration}{
  address={Department of Physics, New Mexico State University, Las Cruces, N.M., U.S.A.}
}

\begin{abstract}
Preliminary results are presented for two analyses of transverse energy ($E_T$) production 
measured with the electromagnetic calorimeters (EMC) of the Pioneering High Energy Nuclear Interaction 
Experiment (PHENIX), in relativistic nuclear interactions in Au+Au heavy-ion collisions
created by the Relativistic Heavy Ion Collider (RHIC), at Brookhaven National Laboratory.
Event-by-event $E_T$ distributions made across collision centrality were used in 
(1) measurements of 200 GeV $\langle E_T \rangle$, and
(2) measurements of 200 GeV and 62.4 GeV $E_T$ distribution relative fluctuations
$\sigma / \langle E_T \rangle$ and $\sigma^2 / \langle E_T \rangle$, where $\sigma$ is the 
standard deviation, and $\sigma^2$ the variance of each semi-inclusive distribution. 
Event centrality was selected in 5\% wide bins and each bin represented by a modeled mean 
number of participant nucleons $\langle N_p \rangle$.
\end{abstract}

\maketitle

\section{Measurements of mean $E_T$ and $E_T$ fluctuations}
The analysis consisted of reconstructing total energy created on each event using 6 
calorimeters centered transverse to the beam and covering pseudorapidity $+/-$ 0.382.
$E_T$ is a multi-particle variable defined as $E_T = \Sigma_i E_i \sin(\theta_i)$, where 
$E_i$ and $\theta_i$ are the energy, and polar angle of the $i^{th}$ particle \cite{Adcox:2001}. 
The $\pi^0$ mass was reconstructed from photon pairs and used to estimate the accuracy of the 
absolute energy scale. Minimum bias events were selected with requirement of at least two
particles incident on each of the two PHENIX Beam-Beam Counters (BBC) placed at forward and 
backward rapidities, and having response correlated to the number of nucleons participating on the event. 
A maximum event vertex of $+/-$ 20 cm relative to the PHENIX origin was accepted. 300,000 events 
were analyzed for $\langle E_T \rangle$, and for fluctuations 8 million 200 GeV events 
and 4 million 62.4 GeV events analyzed. 62.4 GeV event centrality was determined by 
slicing the BBC charge distribution into semi-inclusive distributions of equal numbers of events \cite{Morrison:2004}.
The 200 GeV centrality was determined by slicing a distribution of event BBC charge versus 
energy as measured in the Zero Degree Calorimeters (ZDC), where the response is correlated to the number spectator 
nucleons \cite{Adcox:2001a}. Simulations of the BBC and ZDC responses were used in the definition of 
centrality event classes and related via a Glauber model to the number of participating nucleons on the 
event \cite{Morrison:2004}, \cite{Adcox:2001a}.
The physical variable measured, $\alpha$, describes the functional dependencies $\langle E_T \rangle$ 
\cite{Adler:2005}, and $\sigma / \langle E_T \rangle$ have on event centrality, being of the form
$\langle E_T \rangle \sim N_p^{\alpha}$, and $\sigma / \langle E_T \rangle \sim 1/\sqrt{N_p^{\alpha}}$.
$\alpha$ is obtained from plotting $\langle E_T \rangle$ and 
$\sigma / \langle E_T \rangle$ in centrality, fitting the results with a power law function, and 
then extracting from the fits the powers of $\alpha$.

\section{Measurement errors}
In the fluctuations analyses systematic errors were obtained by measuring 
$\sigma / \langle E_T \rangle$, and $\sigma^2 / \langle E_T \rangle$
for 30 subsets of the data, and the standard deviation of each of the variables calculated across the 
data subsets. This was done for every centrality bin 
and the error in $\sigma / \langle E_T \rangle$ was consistently smaller than the data points except 
in the most peripheral bins (Figures~\ref{var_and_meanET} left, and~\ref{fluctuations}). 
For the 200 GeV mean $E_T$ analysis the systematic errors for each semi-inclusive $\langle E_T \rangle$
measurement were calculated following the procedure outlined in \cite{Adler:2005}.
This systematic error contains a scale component incorporating hadronic 
corrections made to the data which moves the data points together vertically, and a bending 
component which allows for rotation of the data points together about an origin situated at the most 
central semi-inclusive measurement (Fig.~\ref{var_and_meanET}, right), and which is due to the 
limited 93\% efficiency in measuring the total cross section. For both analyses and both data sets 
the statistical error was less than 1\% in each centrality bin.

\section{Results}
\begin{figure}[ht]
\scalebox{.30}{
\includegraphics*[0,0][570,380]{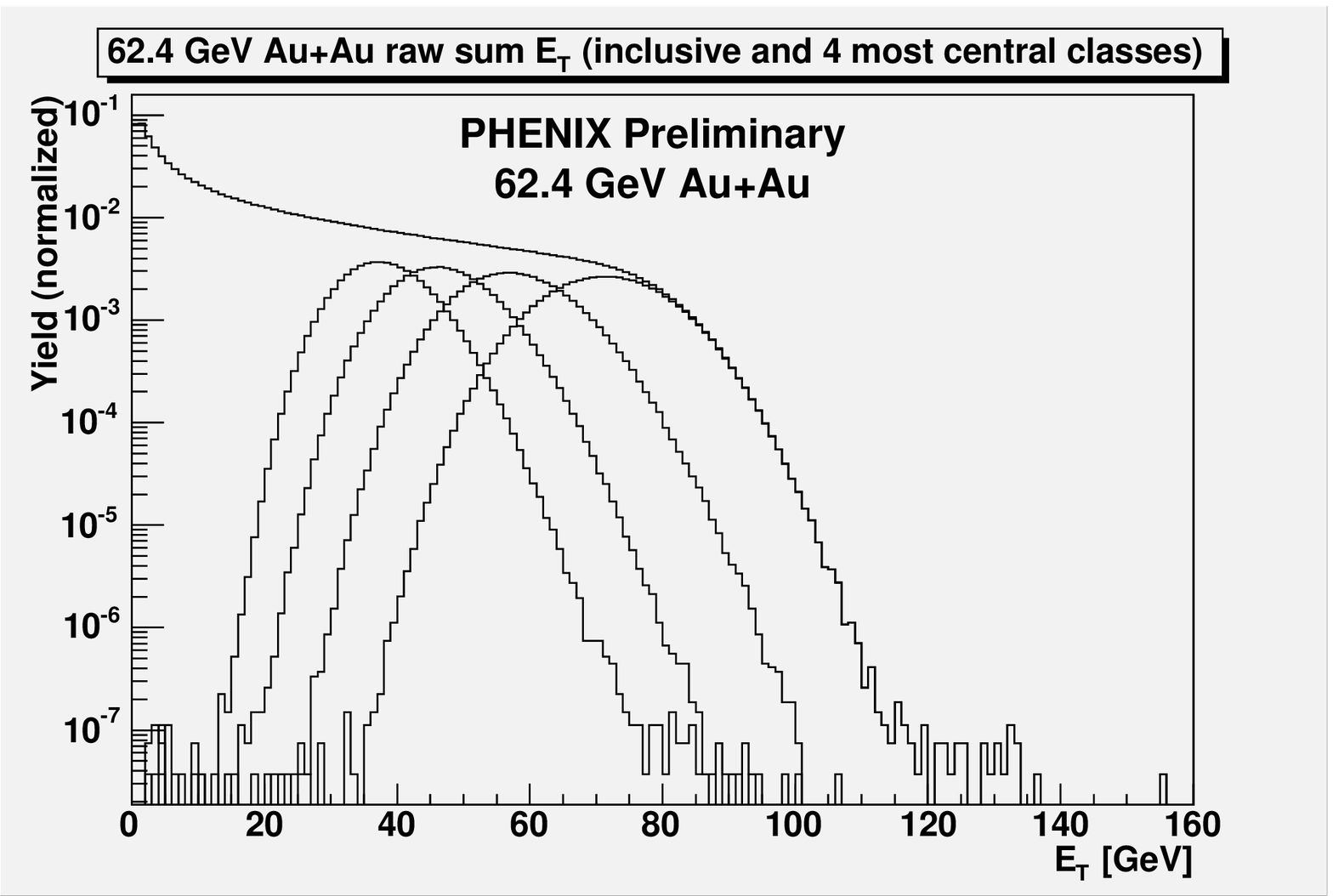}
}
\scalebox{.30}{
\includegraphics*[0,0][570,380]{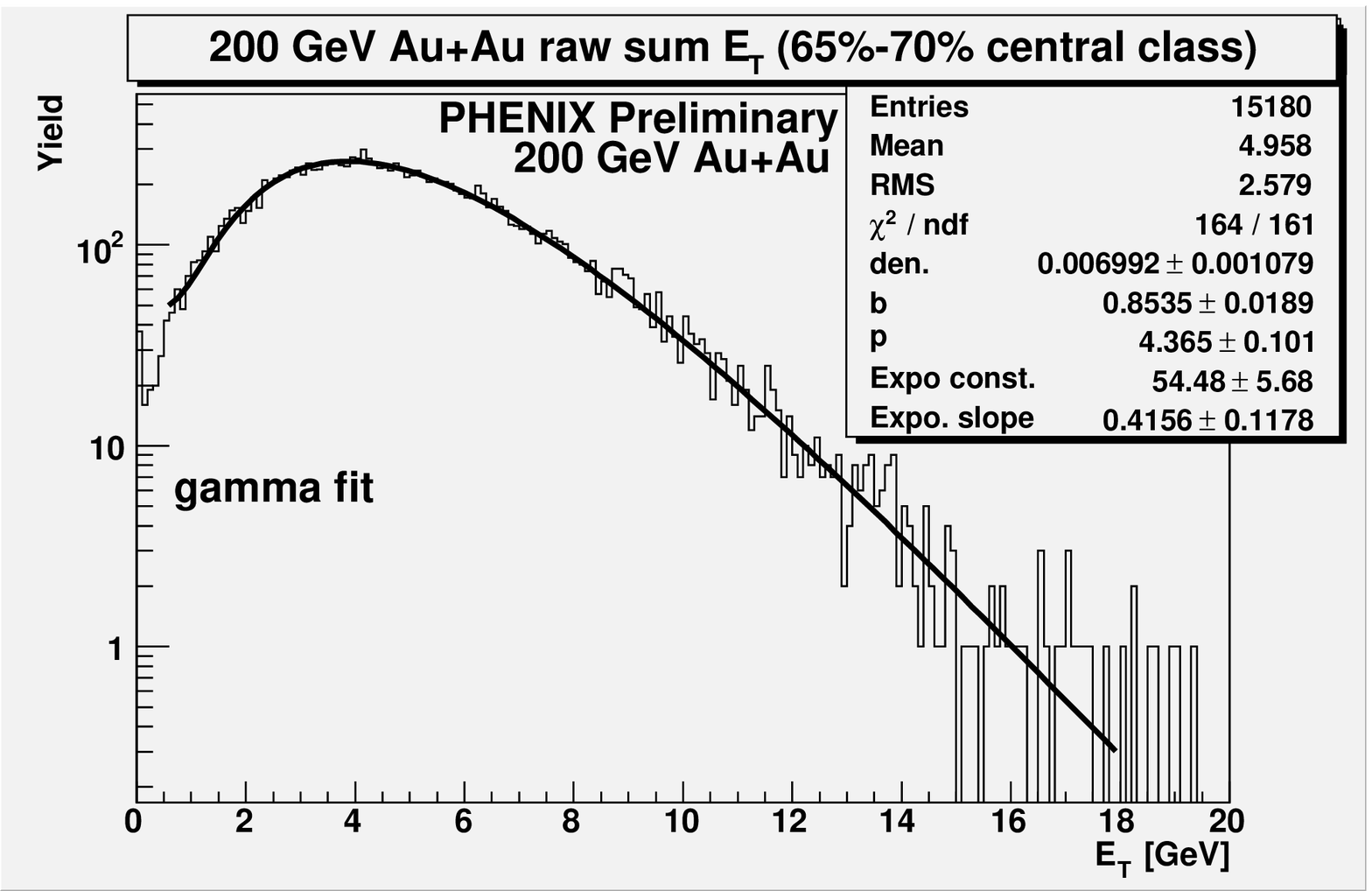}
}
\caption{Event-by-event $E_T$ distributions for (left) 62.4 GeV Au+Au inclusive and four 5\% wide semi-inclusive
classes. (Right) A 200 GeV semi-inclusive distribution fitted to a gamma distribution; gamma distributions
have the property that upon adding them in random combinations, the mean of the resultant distribution is proportional 
to $m$, the number of times the underlying distribution was convolved.}
\label{distributions}
\end{figure}

\begin{figure}[ht]
\scalebox{.295}{
\includegraphics*[0,0][570,380]{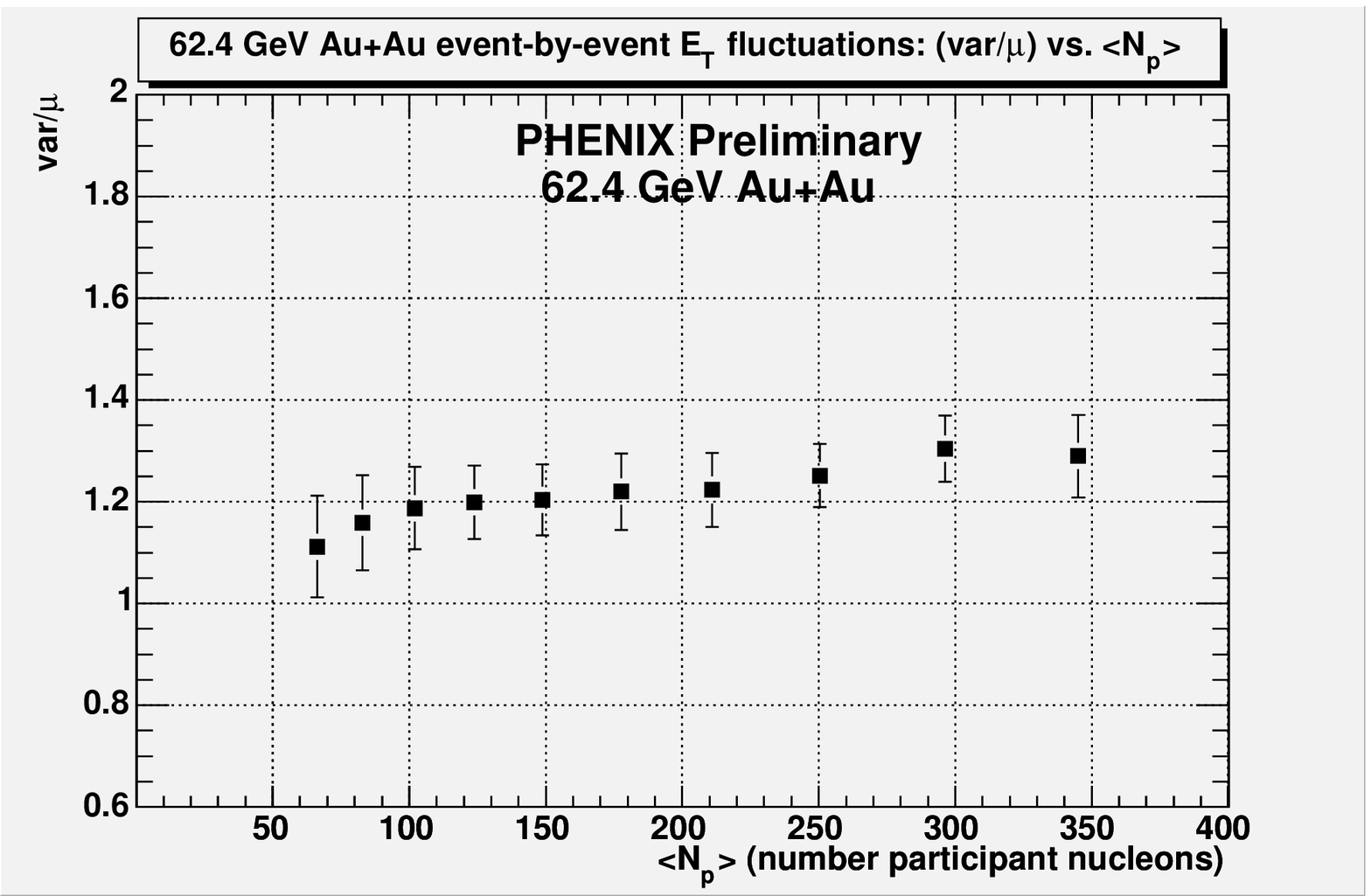}
}
\scalebox{.30}{
\includegraphics*[0,0][570,380]{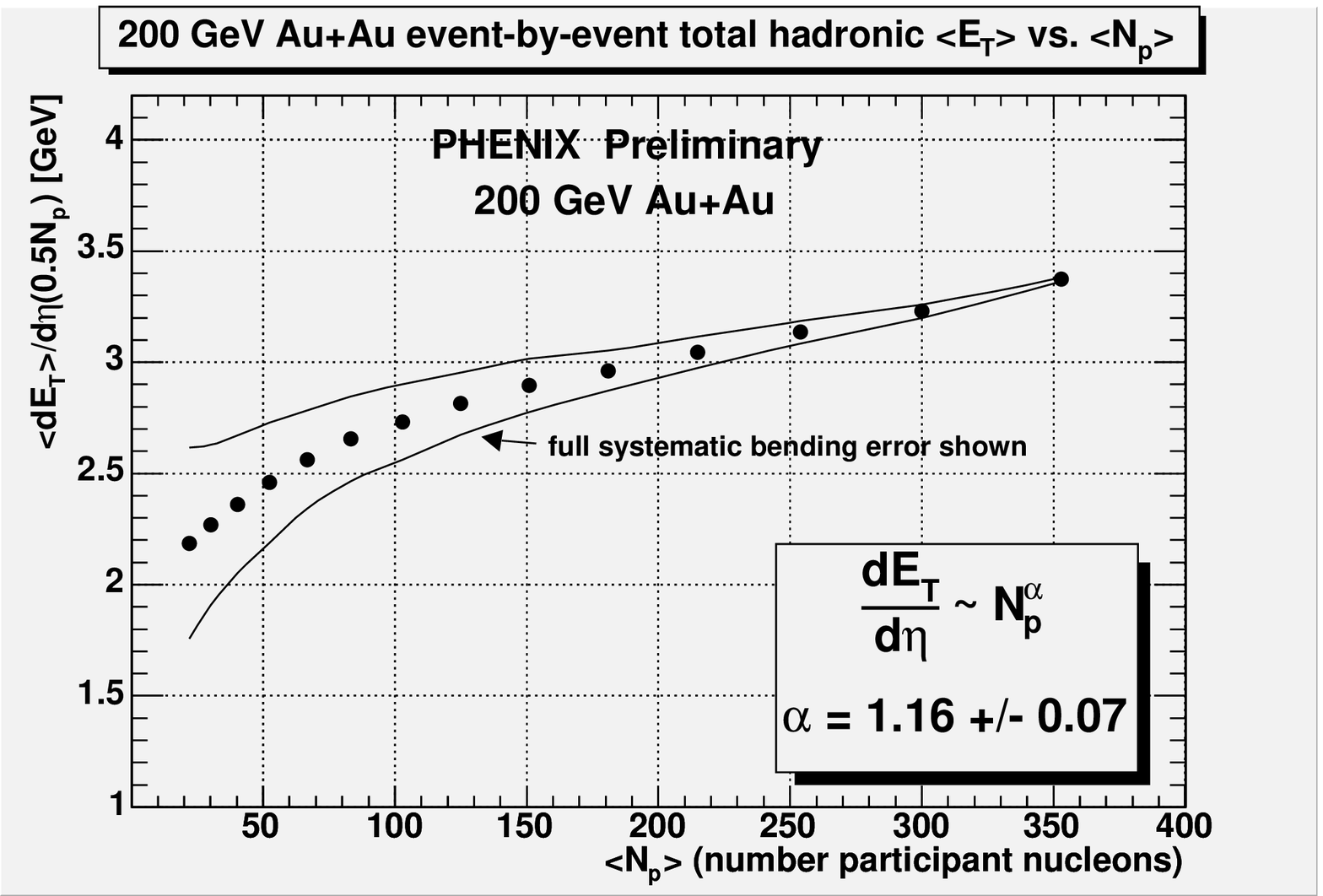}
}
\caption{(Left) 62 GeV $E_T$ distribution relative fluctuation measurements 
$\sigma^2 / \langle E_T \rangle$ ($\sigma^2 \equiv var$, and $\langle E_T \rangle \equiv \mu$) 
versus $\langle N_p \rangle$, shown to be flat within systematic errors; since random 
combinations of a given underlying gamma
distribution preserves $\sigma^2 / \langle E_T \rangle$ this suggests 62.4 GeV 
$E_T$ combines randomly over centrality, the remaining question being what convolves - independent 
collisions, wounded nucleons, participant quarks, ... to be investigated.
(Right) The evolution of 200 GeV Au+Au mean $E_T$ across centrality shown to rise faster than 
participant nucleon linear scaling.}
\label{var_and_meanET}
\end{figure}

\begin{figure}[ht]
\scalebox{.30}{
\includegraphics*[0,0][570,380]{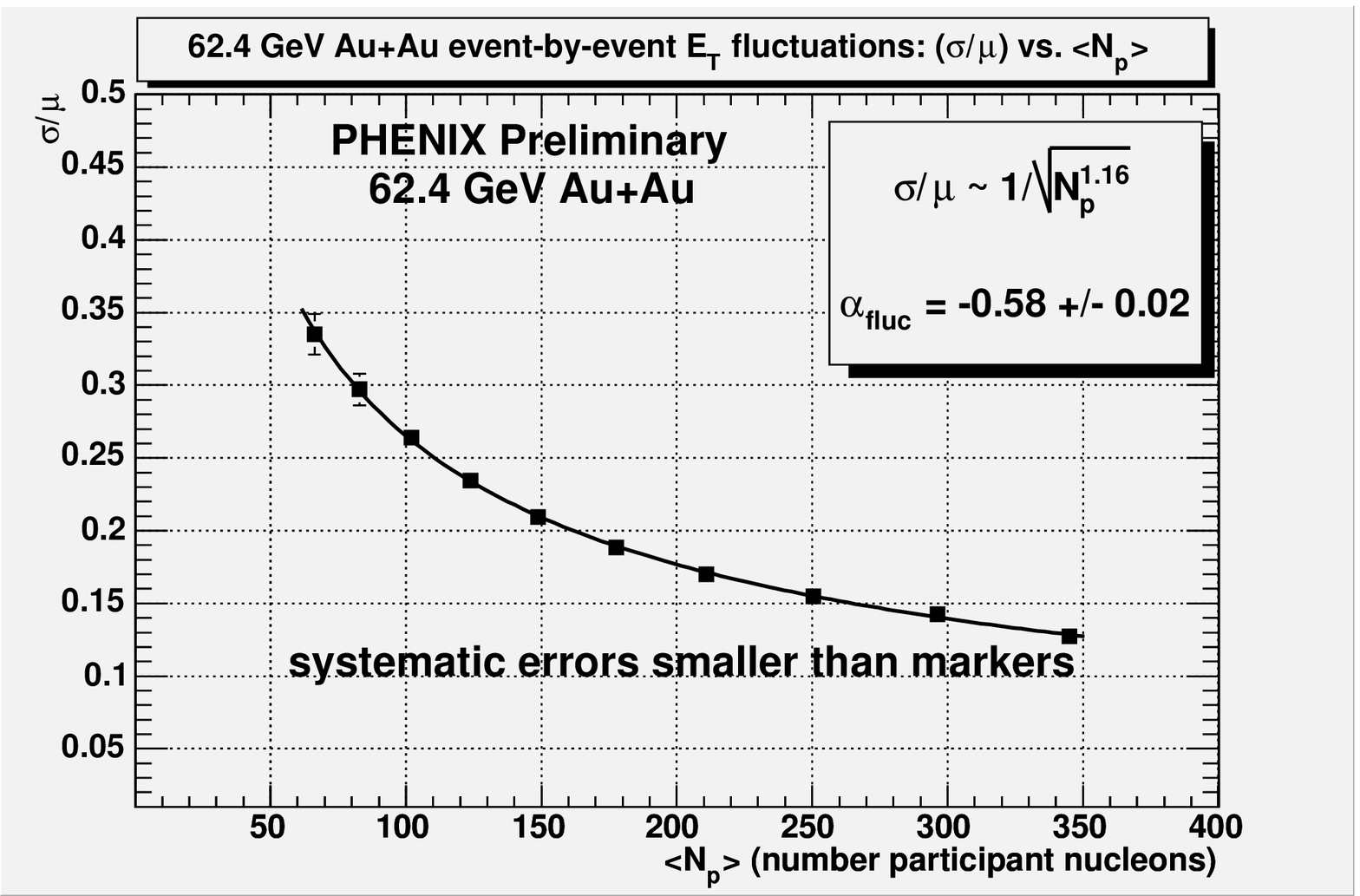}
}
\scalebox{.30}{
\includegraphics*[0,0][570,380]{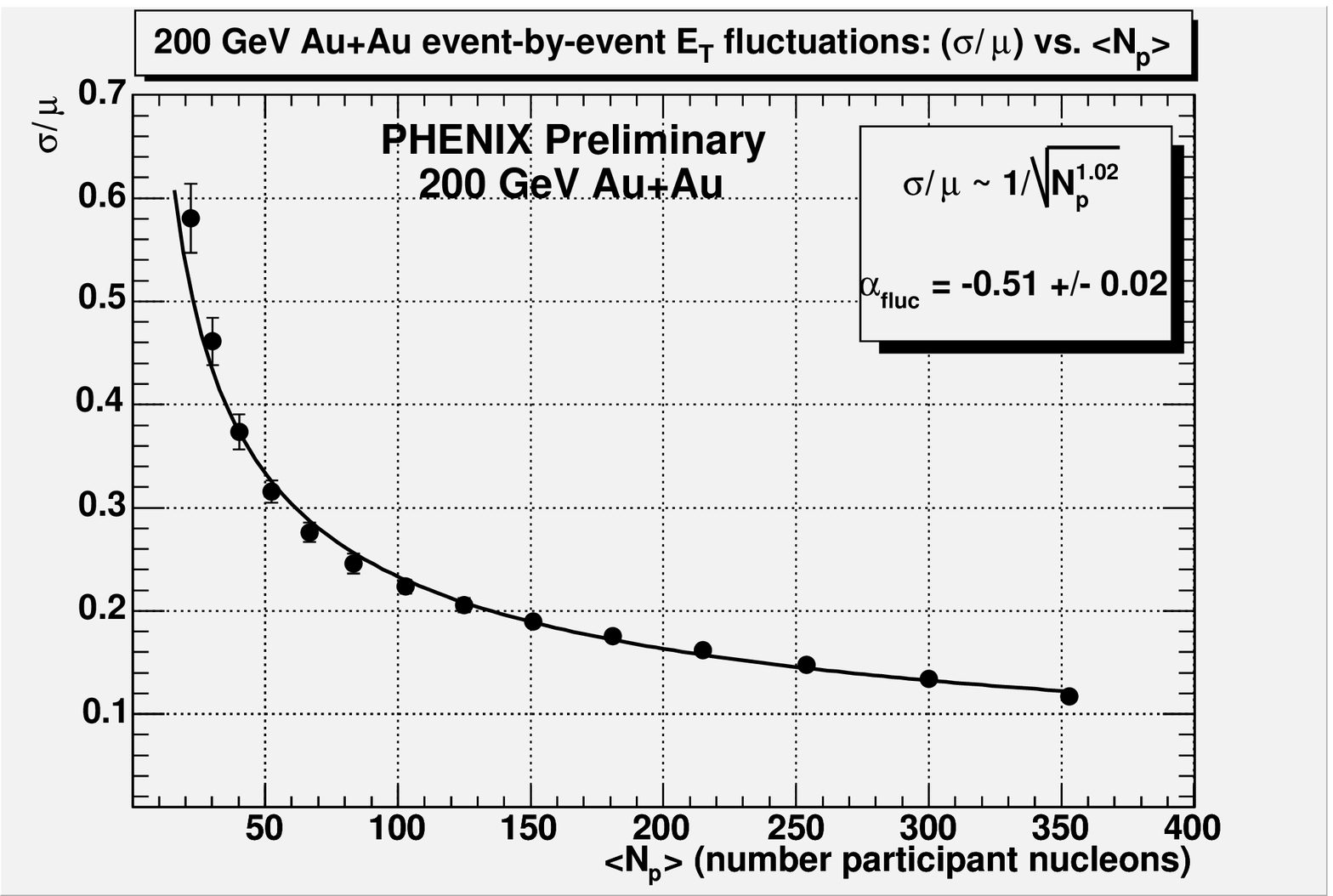}
}
\caption{$\sigma / \langle E_T \rangle$ fluctuations ($\langle E_T \rangle \equiv \mu$) for 
(left) 62 GeV, and (right) 200 GeV distributions. Gamma distributions have the 
property that if just adding random combinations across centrality, the relative fluctuation 
goes as $1/\sqrt{m}$, for the $m^{th}$-fold convolution of 
the underlying distribution. The results for 62 GeV, $N_p^{1.16}$, and 200 GeV, $N_p^{1.02}$, 
were determined by fitting the points to a power law function.}
\label{fluctuations}
\end{figure}

Two results are as follows. 
Firstly, Fig.~\ref{distributions} (right) shows $E_T$ distributions fit a gamma distribution, which has the 
property that in adding random combinations $\sigma^2 / \langle E_T \rangle$ remains flat~\cite{Tannenbaum:2004}. 
Fig.~\ref{var_and_meanET} (left) illustrates that to within the systematic error the 62.4 GeV 
$\sigma^2 / \langle E_T \rangle$ is flat in centrality suggesting that semi-inclusive $E_T$ 
distributions are constructed from the random combinations of an underlying distribution,
the remaining question being what is the underlying mechanism of $E_T$ production at these 
relativistic energies? Secondly, in summing convolutions of gamma distributions,
the mean of the distribution resulting from the $m^{th}$-fold convolution is proportional to $m$, and the fluctuation
$\sigma / \langle E_T \rangle$ proportional to $1/\sqrt{m}$. Figures~\ref{var_and_meanET} (right)
and~\ref{fluctuations} show fit results for the scale dependence $\langle E_T \rangle$, and 
$\sigma / \langle E_T \rangle$ have on $\langle N_p \rangle$; for 200 GeV: 
for ${\langle E_T \rangle}$, $\alpha = 1.16 +/- 0.07$ , and for $\sigma / \langle E_T \rangle$, 
$\alpha = 1.02 +/- 0.04$; for 62.4 GeV $\sigma / \langle E_T \rangle$, $\alpha = 1.16 +/- 0.04$. 
These measurements for the mean, and the fluctuations give approximately the same value of $\alpha$ to within the 
errors, suggesting $E_T$ production measured in the limited aperture solid angle rises 
faster than linear participant nucleon scaling.

\begin{theacknowledgments}
We thank Michael Tannenbaum, Jeffrey Mitchell and Alexander Bazilevsky from Brookhaven National
Laboratory, and Stephen Pate from New Mexico State University.
\end{theacknowledgments}

\bibliographystyle{aipprocl} 
\bibliography{plan_ET}

\clearpage

\end{document}